\documentclass[10pt,jounal,compact,final,comcos]{IEEEtran}

\usepackage{graphicx}

\usepackage{epstopdf}
\DeclareGraphicsExtensions{.pdf,.eps,.png,.jpg,.mps}

\usepackage[cmex10]{amsmath}
\usepackage{algorithm}

\usepackage{algorithmicx}

\usepackage{algpseudocode}

\usepackage{bbm}

\usepackage[tight,footnotesize]{subfigure}

\usepackage{amssymb,amsmath}

\usepackage{epsfig}

\usepackage{stfloats}

\usepackage{endnotes}

\usepackage{multirow}

\usepackage{adjustbox}

\usepackage{subfigure}

\usepackage[subnum]{cases}

\makeatletter
\def\user@resume{resume}
\def\user@intermezzo{intermezzo}
\newcounter{previousequation}
\newcounter{lastsubequation}
\newcounter{savedparentequation}
\makeatother

\newtheorem{lemma}{Lemma}

\newtheorem{theorem}{Theorem}

\begin{document}

%1.\tiny
%2.\scriptsize
%3.\footnotesize
%4.\small
%5.\normalsize
%6.\large
%7.\Large
%8.\LARGE
%9.\huge
%10.\Huge

\title{
\huge
Joint Tx Power Allocation and Rx Power Splitting for SWIPT System with Multiple Nonlinear Energy Harvesting Circuits %\vspace{-0.5em}
}

\author{
Jae-Mo~Kang, Il-Min~Kim,~\IEEEmembership{Senior~Member,~IEEE}, and Dong~In~Kim,~\IEEEmembership{Senior~Member,~IEEE} \vspace{-1.5em}
%\thanks{This work was supported by the Basic Science Research Program through the National Research Foundation of Korea (NRF), funded by the Ministry of Education, Science and Technology (NRF-2013R1A1A2013291).}
%\thanks{The authors are with the Department of Electrical and Computer Engineering, Queen's University, Kingston, ON K7L 3N6, Canada (e-mail: jaemo.kang@queensu.ca; ilmin.kim@queensu.ca).}
\thanks{J.-M. Kang and I.-M. Kim are with the Department of Electrical and Computer Engineering, Queen's University, Kingston, ON K7L 3N6, Canada (e-mail: jaemo.kang@queensu.ca; ilmin.kim@queensu.ca).}
\thanks{D. I. Kim is with the School of Information and Communication Engineering, Sungkyunkwan University (SKKU), Suwon, 440-746, South Korea (e-mail: dikim@skku.ac.kr).}
}

\maketitle

%\vspace{-4em}

\begin{abstract}
%\vspace{-1em}
We study the joint transmit (Tx) power allocation and receive (Rx) power splitting for simultaneous wireless information and power transfer (SWIPT).
Considering the practical scenario of nonlinear energy harvesting (EH), we adopt the realistic nonlinear EH model for analysis.
%For analysis, we adopt the realistic nonlinear EH model.
%Unlike the existing literature that focused on the SWIPT system only with a single EH circuit,
%we consider the SWIPT system with \textit{multiple} EH circuits to effectively address the nonlinearity issue due to the saturation.
To address the critical nonlinearity issue due to the saturation, we propose to use multiple EH circuits in parallel.
%For analysis, we adopt the realistic nonlinear EH model and the most general dynamic power splitting scheme.
%To characterize the optimal R-E tradeoff for the nonlinear EH with multiple EH circuits, the problem is formulated to maximize the achievable rate under the harvested energy constraint and the average power constraint,
%The problem is formulated to maximize the achievable rate under the harvested energy constraint and the average power constraint,
%which is nonconvex.
%Solving this challenging problem, we develop the jointly optimal Tx power allocation and Rx power splitting scheme.
An important problem is to maximize the achievable rate by jointly optimizing Tx power allocation and Rx power splitting, which is a nonconvex problem.
In this paper, we first derive the optimal solution for any number of EH circuits.
Then we study how the number of EH circuits required to avoid the saturation should be determined.
%Also, as a special case of this scheme, we present the jointly optimal Tx power allocation and Rx time switching scheme.
%Then, for these two schemes, we determine the number of EH circuits required to achieve the same R-E tradeoff performance as the ideal linear EH model.
From the obtained results, we draw useful and interesting insights into the SWIPT system with nonlinear EH.
Numerical results demonstrate that employing multiple EH circuits substantially enhances the SWIPT performance with nonlinear EH.
\end{abstract}

%\vspace{-0.5em}

\begin{IEEEkeywords}
%\vspace{-1em}
%Multiple energy harvesting circuits, nonlinear energy harvesting, power splitting, rate-energy (R-E) tradeoff, simultaneous wireless information and power transfer (SWIPT).
Multiple energy harvesting circuits, nonlinear energy harvesting, power allocation, power splitting, SWIPT.
\end{IEEEkeywords}

\IEEEpeerreviewmaketitle

%\vspace{-1.5em}
\section{Introduction}
%\vspace{-0.5em}

Simultaneous wireless information and power transfer (SWIPT) using radio frequency (RF) signals
has been extensively studied in the literature \cite{Zhou13}--\cite{Shi}.
In the most existing works on the SWIPT including \cite{Zhou13}--\cite{Kim16},
it was assumed that the amount of harvested energy linearly increases indefinitely with the input RF power of the energy harvesting (EH) circuit, namely, the {\it linear} EH model.
However, this assumption is too idealistic because the linearity is valid only when the energy conversion efficiency is constant over the infinitely wide range of the input power level.
As validated in many experimental results \cite{Valenta14}, \cite{Guo12}, the practical EH circuit exhibits the nonlinear behavior because the energy conversion efficiency is different (not constant) depending on the input power level.
%Specifically, only when the input power is below a certain level, the energy conversion efficiency is not small and the amount of harvested energy increases almost linearly.
%On the other hand, when the input power exceeds a certain level, the energy conversion efficiency is very small (close to zero) and the amount of harvested energy saturates.

Very recently, to overcome the critical limitations of the linear EH model and to address the practicality issue of nonlinear EH,
the SWIPT was studied for the realistic nonlinear EH models \cite{Kang}--\cite{Shi}.
Among the various nonlinearity issues for EH, the most critical issue for the performance is the nonlinearity due to the saturation,
because it severely limits the amount of harvested energy and the energy conversion efficiency of the EH circuit.
Therefore, overcoming the nonlinearity due to the saturation is a practically very important issue in the SWIPT system.
In the previous works on the nonlinear EH \cite{Kang}--\cite{Shi}, various approaches have been developed to cope with the nonlinearity.
However, none of these approaches were effective to overcome the saturation nonlinearity,
because the approaches were developed only with a single (nonlinear) EH circuit. Once the EH circuit saturates, there is no further performance improvement in the amount of harvested energy. An obvious, yet effective, way to overcome this limitation is to use multiple EH circuits in parallel.
Then a very fundamental and important (but, non-trivial) question is: How to optimize the entire SWIPT system?; more specifically, how to jointly optimize the transmit (Tx) power and the receive (Rx) power splitting ratio? and how to determine the number of EH circuits that must be turned on?
In the literature, this fundamental issue has not been studied. This motivated our work.

%Then a very fundamental and important (but, non-trivial) question is: what is the ultimate performance limit for the SWIPT system with multiple nonlinear EH circuits? In the literature including \cite{Kang}--\cite{Shi}, this fundamental question has not been answered. This motivated our work.
%and how many nonlinear EH circuits do we need to achieve the same performance as the linear EH circuit?

In this paper, we study the joint Tx power allocation and Rx power splitting for the SWIPT system with multiple nonlinear EH circuits.
Adopting a realistic nonlinear EH model, we formulate the optimization problem to maximize the achievable rate with the harvested energy constraint and the average power constraint,
which is nonconvex, and thus, challenging to solve.
The contributions of this paper are as follows.
First, for any number of EH circuits, we develop the jointly optimal Tx power allocation and Rx power splitting scheme.
Second, we determine the required number of EH circuits to overcome the saturation nonlinearity.
%As a special case of this scheme, we present the jointly optimal Tx power allocation and Rx \textit{time switching} scheme.
Also, from the obtained results, we draw various interesting and useful insights into the SWIPT system with nonlinear EH.

%This paper is organized as follows.
%In Section II, the system model is described and the proposed SWIPT schemes are presented.
%In Section III, the analysis is conducted for the case of fixed transmit power.
%In section IV, the analysis is extended to the case of transmit power allocation.
%In section V, the applicability of the proposed SWIPT schemes is discussed.
%Section VI concludes this paper.

%\textit{Notation:} We use $\Omega^{C}$ and $| \Omega |$ to denote the complement and the cardinality of a set $\Omega$, respectively.
%Also, $\lceil m \rceil$ denotes the smallest integer not less than $m$.

%\vspace{-1.em}
\section{System Model and Problem Formulation}
%\vspace{-0.5em}

We consider a point-to-point SWIPT system with one transmitter and one receiver, each equipped with a single antenna.
Each block consists of $N$ transmitted symbols:
at the $k$th symbol period, the symbol is transmitted with power $P_{k} \geq 0$, where $k \in \{ 1, \cdots, N \}$.
The number $N$ of symbols is assumed to be sufficiently large such that $\alpha N$ is an integer for arbitrary $0 < \alpha < 1 $.
Let $h$ denote the power gain of the channel between the transmitter and the receiver,
which is assumed to be quasi-static.
%\footnote{
%The quasi-static assumption is quite reasonable in the most SWIPT applications
%since the current SWIPT is practically feasible only in the stationary environment with low mobility, e.g., indoor scenario over the fixed range of distances (3--15 m) \cite{Lu_survey}, \cite{Kim16_2}.
%}
%i.e., it remains constant over the time duration of the block.
Also, the channel state information (CSI) is assumed to be known at the transmitter and the receiver.\footnote{Recently, the received RF power based channel estimation scheme has been developed in \cite{Choi}, which can be used in our system to acquire the CSI at the transmitter and the receiver.
}

%\vspace{-1.em}
\subsection{Linear and Nonlinear Energy Harvesting}
%\vspace{-0.5em}

In the previous works including \cite{Zhou13}--\cite{Kim16}, the linear EH model was adopted.
In the linear EH model, the amount of harvested energy $Q_{\texttt{L}}$ over the time duration of $T$ is linearly proportional to the input power $P_{\rm in}$ of the EH circuit as follows: 
%$Q_{\texttt{L}}   =   \zeta P_{\rm in} T$,
\begin{align}
\label{Q_L}
Q_{\texttt{L}}   =   \zeta P_{\rm in} T
\end{align}
where $ 0 <   \zeta  \leq 1 $ is the energy conversion efficiency of the EH circuit, which is assumed be a constant independent of the input power.
However, as validated in the experimental results \cite{Valenta14}, \cite{Guo12} and as analyzed in \cite{Clerckx},
the energy conversion efficiency of the actual EH circuit is different (not constant) over the different input power levels, meaning that the amount of harvested energy increases nonlinearly with the input power.
%Specifically, when the input power is below a certain level (i.e., the rectifier is turned-on and operates in the linear region),
%the energy conversion efficiency is large (about 0.7 at the frequency of 915 MHz) and, in this range of input power, the amount of harvested energy increases almost linearly with the input power (e.g., see \cite[Fig. 2]{Boshkovska15}, \cite[Fig. 2]{Shi}).
Specifically, \textit{only} when the input power is below a certain level, the energy conversion efficiency is not small (about 0.7 at the frequency of 915 MHz), and the amount of harvested energy increases almost linearly with the input power (e.g., see \cite[Fig. 2]{Boshkovska15}, \cite[Fig. 2]{Shi}).
On the other hand, when the input power exceeds a certain level,
the energy conversion efficiency becomes very small (close to zero) due to the reverse breakdown, and the amount of harvested energy saturates.
%Unfortunately, the ideal and simplistic linear model cannot accurately model the nonlinear behavior of the actual EH circuit (particularly, the nonlinearity due to the saturation).
%Thus, the use of the linear model may incur severe mismatch or inaccuracy in the practical system.

In order to accurately model the nonlinear behavior of the practical EH circuit,
%and to overcome the critical limits of the linear EH model,
several realistic nonlinear EH models have been suggested and studied in the recent literature \cite{Kang}--\cite{Shi}.
Among the various nonlinear EH models, the nonlinear model used in \cite{Dong16}, \cite{Shi} is mathematically tractable
and is shown to accurately match the experimental results \cite[Fig. 2]{Shi}.
In this paper, for accuracy, practicality, and tractability of the analysis with useful insights,
we adopt the nonlinear model of \cite{Dong16}, \cite{Shi}.
%One may extend the analysis presented in this paper to the other nonlinear models considered in \cite{Clerckx}--\cite{Boshkovska17}.
In this nonlinear model, the amount of harvested energy is modeled based on the piecewise linear function as follows:
\begin{align}
\label{Q_NL}
Q_{\texttt{NL}} & = \begin{cases}  \zeta P_{\rm in} T , & {\rm if~} \zeta P_{\rm in}  \leq P_{s}  \\  P_{s} T ,  &  {\rm if~} \zeta P_{\rm in} > P_{s} \end{cases}
\end{align}
where
$ P_{s}$ $(\leq \zeta P_{\rm in})$ denotes the maximum harvested power when the EH circuit is saturated.\footnote{
For a single diode rectifier, the maximum harvested power is given by $ P_{s} = \frac{ v_{b}^{2} }{ 4 r_{l} }$,
where $v_{b}$ is the reverse breakdown voltage of the diode and $r_{l}$ is the resistance of the load \cite{Valenta14}.
}
%Due to the law of energy conservation, it follows that $P_{s} \leq \zeta h P$.
%Since the maximum harvested energy should be limited by that by the linear EH model, we have $E_{s} \leq \zeta h P$.
%Given the measurement data, the parameters $P_{s}$ and $\zeta$ can be determined by the curve fitting.
%Note that when $P_{s} \rightarrow \infty$, i.e., there is no saturation,
%the nonlinear EH model $Q_{\texttt{NL}}$ of (\ref{Q_NL}) approaches the linear EH model $Q_{\texttt{L}}$ of (\ref{Q_L}).
%In this sense, therefore, the linear model can be considered as a special case of the adopted nonlinear model.

%\vspace{-1.em}
\subsection{SWIPT with Multiple Nonlinear EH Circuits}
%\vspace{-0.5em}

In this paper, for analysis, we consider the dynamic power splitting architecture \cite[Sec. III-A]{Zhou13}, which is the most general architecture for the SWIPT.
At the receiver, the received power at the $k$th symbol period is dynamically split with a power splitting ratio $0  \leq  \rho_{k}  \leq  1$.
At the transmitter, the transmit power $P_{k}$ is dynamically adjusted under the average power constraint $\frac{1}{N} \sum_{k=1}^{N} P_{k} \leq P$,
where $P$ denotes a threshold for the average transmit power.

First, the $(1 - \rho_{k})$ portion of the received power, i.e., $(1 - \rho_{k}) h P_{k}$, is used for information decoding (ID). The average achievable rate is given by
\begin{align}
\label{sum_capacity}
\mathcal{R} ( \mathbf{P}, \boldsymbol{\rho} )  =  \frac{1}{N}  \sum_{k=1}^{N} \mathbb{R} ( P_{k}, \rho_{k} ) = \frac{1}{N}  \sum_{k \in \Omega} \mathbb{R} ( P_{k}, \rho_{k} )
\end{align}
where $ \mathbf{P} $ and $ \boldsymbol{\rho} $ are the vectors composed of $P_{k}$'s and $\rho_{k}$'s, respectively.
Also, $ \Omega = \{ k : 0 \leq \rho_{k} < 1 \}$ and
\begin{align}
\label{capacity}
\mathbb{R}   ( x, y )  =  \log_{2} \left( 1  +  \frac{ (1 - y ) h x }{ ( 1 - y) \sigma_{ {\rm A}}^{2} + \sigma_{\rm cov}^{2} }  \right).
\end{align}
In (\ref{capacity}), $ \sigma_{ {\rm A}}^{2} $ and $\sigma_{\rm cov}^{2}$ are the variances of the antenna noise and the RF-to-baseband conversion noise, respectively.

Second, the remaining $\rho_{k}$ portion of the received power is used for EH, and thus, the input power of the EH circuit is given by $P_{\rm in} = \rho_{k} h P_{k}$.
For the case of nonlinear EH, the amount of harvested energy as well as the energy conversion efficiency is strictly limited by the saturation effect,
which is a critical issue for the performance of the SWIPT system. To address this issue, we propose to use multiple (nonlinear) EH circuits.
Specifically, the input power is evenly split among $M$ $(\geq 1)$ EH circuits,
%such that the probability that each circuit will operate in the saturation region is reduced.
such that no EH circuit enters the saturation region.
Taking this approach and using the realistic nonlinear EH model of (\ref{Q_NL}), the average net harvested energy can be written as
\begin{align}
\label{sum_energy}
\mathcal{Q}_{\texttt{NL}}  ( \mathbf{P}, \boldsymbol{\rho} )  =  \frac{1}{N} & \Bigg[  \sum_{k \in \Omega^{C} } \sum_{i=1}^{ M } \mathbb{Q}_{\texttt{NL}} \left( P_{k}, \frac{1}{M} \right) \nonumber\\
&  +  \sum_{k \in \Omega } \left( \sum_{i=1}^{ M } \mathbb{Q}_{\texttt{NL}} \left( P_{k}, \frac{\rho_{k}}{M} \right) - P_{c} T \right) \Bigg]
\end{align}
where
$ \Omega^{C} = \{ k : \rho_{k} = 1 \}$ is the complement of the set $\Omega$.
Also, $P_{c} $ $ ( < P_{s} )$ denotes the circuit power consumed by the ID circuitry\footnote{
Only the power consumption by ID circuity is considered, because no power is consumed by the EH circuitry which consists of the passive devices such as the diode, inductor, and capacitor \cite{Zhou13}.
}
and
\begin{align}
\label{E_NL}
\mathbb{Q}_{\texttt{NL}} ( x, y ) & =  \begin{cases} \zeta x y h T , & {\rm if~} \zeta x y h \leq P_{s}  \\  P_{s} T ,  &  {\rm if~} \zeta x y h > P_{s} \end{cases}.
\end{align}

%From (\ref{sum_capacity}) and (\ref{sum_energy}), we can define the R-E region as follows:
%\begin{align}
%\label{RE_region}
%\mathcal{C}_{\texttt{NL}}^{M} = \bigcup_{ \mathbf{P}, \boldsymbol{\rho} } \Big\{  (R,Q):   R  \leq  \mathcal{R} ( \mathbf{P}, \boldsymbol{\rho} ) , Q  \leq  \mathcal{Q}_{\texttt{NL}}  ( \mathbf{P}, \boldsymbol{\rho} ) \Big\},
%\end{align}
%which contains all possible pairs of the rate and harvested energy for any number $M$ of EH circuits.

%\vspace{-1.em}
\subsection{Problem Formulation}
%\vspace{-0.5em}

%In this paper, our goal is to maximize the R-E region $\mathcal{C}_{\texttt{NL}}^{M}$ of (\ref{RE_region})
%to characterize the optimal R-E tradeoff of the SWIPT system with multiple EH circuits.
%To this end, we formulate the optimization problem as follows:
In this paper, using the realistic nonlinear EH model, we aim to develop the jointly optimal Tx power allocation and Rx power splitting scheme for the SWIPT system with any number $M$ of EH circuits,
in the sense of maximizing the achievable rate under the constraints on the harvested energy and the average power.
Thus, the problem is formulated as follows:
%\begin{subequations}
%\label{P1}
\begin{align}
\label{P1_obj}
{\rm (P1):} \quad \underset{ \mathbf{P}, \boldsymbol{\rho}  } {\max} & \quad  \mathcal{R} ( \mathbf{P}, \boldsymbol{\rho} )  \\
\label{P1_const_1}
{\rm s.t.} & \quad \mathcal{Q}_{\texttt{NL}}  ( \mathbf{P}, \boldsymbol{\rho} ) \geq Q ,  \quad \frac{1}{N} \sum_{k=1}^{N} P_{k} \leq P 
%\label{P1_const_2}
%& \quad  \frac{1}{N} \sum_{k=1}^{N} P_{k} \leq P, \\
%\label{P1_const_3}
%& \quad  P_{k} \geq 0, ~ 0 \leq \rho_{k} \leq 1, ~ \forall k
\end{align}
%\end{subequations}
where $Q$ is a threshold for the harvested energy.
%In (P1), since the constraints of (\ref{P1_const_1}) and (\ref{P1_const_2}) must be satisfied with equalities at the optimal point,
%the largest R-E region, i.e., the optimal R-E tradeoff, can be achieved by solving (P1) for all possible values of $Q \in \left[ 0 , Q_{\max} \right]$, where $Q_{\max} = M \mathcal{Q}_{\texttt{NL}} \left( P, \frac{1}{M} \right)$. However, it is generally very challenging to solve (P1) due to the nonconvexity.
The problem (P1) is generally very challenging to solve due to the nonconvexity.
%In \cite[eq. (11)]{Zhang13}, an optimization problem similar to (P1) was studied for the linear EH model.
%However, in \cite{Zhang13}, the solution was derived only for a special case when $\rho_{k} \in \{ 0, 1 \}$, $\forall k$, $M = 1$, and $P_{c} = 0$.
To the best of our knowledge, in the literature, (P1) still remains unsolved even for the {\it linear} EH model,\footnote{
In \cite[eq. (11)]{Zhang13}, an optimization problem similar to (P1) was studied for the linear EH model.
However, in \cite{Zhang13}, the solution was derived only for a special case of $\rho_{k} \in \{ 0, 1 \}$, $\forall k$, and $P_{c} = 0$,
i.e., the time switching architecture \cite[eq. (15)]{Zhou13}, with $M = 1$.
}
not to mention the {\it nonlinear} EH model.
%Furthermore, the solution of \cite{Zhang13}, \cite{X_Zhou_2} is no longer optimal to (P1) because the linear and nonlinear EH models are mathematically and practically different.

%\vspace{-1.em}
%\section{Joint Tx Power Allocation and Rx Power Splitting for Nonlinear EH with Multiple EH Circuits}
\section{Joint Tx Power Allocation and Rx Power Splitting with Multiple Nonlinear EH Circuits}
%\vspace{-0.5em}

\subsection{Optimal Solution to (P1)}
%\vspace{-0.5em}

In this subsection, we derive the optimal solution to (P1) by converting it into a more tractable form. To this end, in the following, we first derive the optimal structure of the dynamic power splitting scheme.
\begin{lemma}
\label{lem_1}
The solution to (P1) takes the following form:
%\begin{subequations}
\begin{align}
\label{P_k}
P_{k} & =  \begin{cases}  P_{\rm EH}, & k=1,\cdots,\alpha N \\   P_{\rm ID}, & k= \alpha N + 1, \cdots, N \end{cases}, \\
\label{rho_k}
\rho_{k} & = \begin{cases} 1,  & k=1,\cdots,\alpha N \\  \rho , & k= \alpha N + 1, \cdots, N \end{cases}
\end{align}
%\end{subequations}
where $P_{\rm EH} \geq 0 $, $P_{\rm ID} \geq 0 $, $ 0 \leq \alpha \leq 1$, and $ 0 \leq \rho < 1 $ are the variables to be determined.
\end{lemma}
\begin{IEEEproof}
See Appendix A.
\end{IEEEproof}

Lemma \ref{lem_1} means that only the EH (no ID) has to be carried out during the $\alpha$ portion of the block,
and both EH and ID has to be carried out during the remaining $(1 - \alpha)$ portion of the block.
The result of Lemma \ref{lem_1} is very interesting and practically useful, because it indicates that the jointly optimal Tx power allocation and Rx power splitting scheme reduces to the on-off power splitting architecture \cite[eq. (17)]{Zhou13}, which is much simpler than the dynamic power splitting architecture.
Furthermore, it is very important to note that using Lemma \ref{lem_1}, the original optimization of (P1) over the two $N$-dimensional vectors $\mathbf{P}$ and $\boldsymbol{\rho}$ can be substantially simplified to the optimization only over the four scalars $P_{\rm EH}$, $P_{\rm ID}$, $\rho$, and $\alpha$, as follows:
%\begin{subequations}
%\label{P1_}
\begin{align}
\label{P1_obj_}
{\rm (P1'):} & \quad \underset{  P_{\rm EH}, P_{\rm ID}, \alpha, \rho } {\max}  \quad (1 - \alpha) \mathbb{R} ( P_{\rm ID}, \rho )   \\
\label{P1_const_1_}
{\rm s.t.} & \quad \alpha M \mathbb{Q}_{\texttt{NL}} \left( P_{\rm EH}, \frac{1}{M} \right) + (1 - \alpha) M \mathbb{Q}_{\texttt{NL}} \left( P_{\rm ID}, \frac{\rho}{M} \right) \nonumber\\
& \quad  - (1 -\alpha) P_{c} T \geq Q,  \\
\label{P1_const_2_}
& \quad \alpha P_{\rm EH} + (1 - \alpha) P_{\rm ID} \leq P 
%\label{P1_const_3_}
%& \quad P_{\rm EH} \geq 0, ~ P_{\rm ID} \geq 0, ~ 0 \leq \alpha \leq 1, ~ 0 \leq \rho < 1.
\end{align}
%\end{subequations}
Note that by Lemma \ref{lem_1}, it possible to significantly reduce the complexity to solve (P1).
However, the converted problem (P1$'$) is still nonconvex because the variables are coupled.
In this paper, by determining the variables $P_{\rm EH}$, $P_{\rm ID}$, and $\rho$ in terms of $\alpha$,
we can obtain the optimal solution to (P1$'$) very efficiently, as shown in the following.
\begin{theorem}
\label{thm_1}
The solution to (P1$'$) is given by
\begin{align}
\label{P1_sol_alpha}
& \alpha^{*}  = \arg \underset{ \alpha_{\rm low} \leq \alpha \leq 1 }{\max} ( 1 - \alpha) \mathbb{R} \big( P_{\rm ID} (\alpha) , \rho(\alpha) \big), \\
\label{P1_sol_rho}
& \rho^{*} ( \alpha^{*} )   =  \min  \left\{ \rho_{1} ( \alpha^{*} ) , \rho_{2} ( \alpha^{*} )  \right\}, \\
\label{P1_sol_pe}
& P_{\rm EH}^{*} ( \alpha^{*}  )  =  \begin{cases} \frac{Q + (1 - \alpha^{*}) P_{c} T - \zeta \rho^{*} ( \alpha^{*} ) h P T}{ \alpha^{*} \zeta \left( 1 - \rho^{*} ( \alpha^{*} ) \right)  h T } ,  & {\rm if ~}  \alpha^{*} > 0 \\ 0 , & {\rm if ~}  \alpha^{*} = 0  \end{cases},  \\
\label{P1_sol_pi}
& P_{\rm ID}^{*} ( \alpha^{*}  )  =  \begin{cases} \frac{\zeta h P T - Q - (1 - \alpha^{*}) P_{c} T}{ (1 -  \alpha^{*} ) \zeta \left( 1 - \rho^{*}( \alpha^{*} ) \right)  h T} ,  & {\rm if ~}  \alpha^{*} < 1  \\ 0 , & {\rm if ~}  \alpha^{*} = 1 \end{cases}  .
\end{align}
In (\ref{P1_sol_alpha}), $ \alpha_{\rm low} = \max \left\{  1 -  \frac{ \zeta h P T - Q }{P_{c} T}  , 0 \right\} $. Also, $\rho (\alpha)$ and $P_{\rm ID} (\alpha)$ are defined similarly as in (\ref{P1_sol_alpha}) and (\ref{P1_sol_pi}), respectively. In (\ref{P1_sol_alpha}), the value of $\alpha^{*}$ can be determined by the one-dimensional searching.
In (\ref{P1_sol_rho}), $\rho_{1} ( \alpha ) = \frac{ Q + (1 - \alpha) P_{c} T }{ \zeta h P T }$ and $\rho_{2} ( \alpha ) = \frac{ (1 - \alpha) M P_{s} T }{ (1 - \alpha) M P_{s} T + \zeta h P T - Q - (1 - \alpha) P_{c} T }$.
\end{theorem}
\begin{IEEEproof}
See Appendix B.
\end{IEEEproof}

Substituting (\ref{P1_sol_alpha})--(\ref{P1_sol_pi}) into (\ref{P_k}) and (\ref{rho_k}), the optimal solution to (P1) can be obtained.
The result of Theorem \ref{thm_1} is very useful in practice due to its very low complexity. From Theorem \ref{thm_1}, one can also obtain the insights as follows:
As $\alpha^{*}$ decreases (except $\alpha^{*} = 0$), the power $P_{\rm EH}^{*} ( \alpha^{*}  )$ used only for EH increases to meet the harvested energy constraint.
The remaining power $P_{\rm ID}^{*} ( \alpha^{*}  )$ is used for both EH and ID, and thus, it decreases as $\alpha^{*}$ decreases.

\subsection{Determining the Number of EH Circuits}
%\vspace{-0.5em}

In the previous subsection, we derived the optimal solution to (P1) for any given number $M$ $(\geq 1)$ of EH circuits.
In this subsection, we determine the number of EH circuits required to overcome the saturation nonlinearity.
The fundamental idea is as follows: The number $M$ of EH circuits increases one by one until none of the EH circuits operate in the saturation region.
Specifically, initially setting $M = 1$, the value of $M$ increases to $M+1$ if the (effective) input power $ P_{\rm in}^{M} = \max \left\{ h P_{\rm EH}^{*} ( \alpha^{*}  ) , h P_{\rm ID}^{*} ( \alpha^{*}  ) \right\} -  \frac{(M-1) P_{s}}{\zeta} $ fed into the $M$th circuit exceeds the saturation threshold $\frac{P_{s}}{\zeta}$.
This proceeds until $M$ reaches $M_{\rm max}$, where $M_{\rm max}$ is the maximum number of EH circuits.
The proposed algorithm is presented in Algorithm \ref{A1}.
\begin{algorithm}[h]
  \small
  \caption{\small Proposed algorithm for determining the number of EH circuits}
  \label{A1}
  \begin{algorithmic}[1]
          \State Set $M=1$.
          \While{$M > M_{\max}$}
            \State Compute $P_{\rm EH}^{*} ( \alpha^{*} )$ and $P_{\rm ID}^{*} ( \alpha^{*} )$ according to (\ref{P1_sol_pe}) and (\ref{P1_sol_pi}), respectively.
            \State Compute $ P_{\rm in}^{M} =  \max \left\{ h P_{\rm EH}^{*} ( \alpha^{*}  ) , h P_{\rm ID}^{*} ( \alpha^{*}  ) \right\} -  \frac{(M-1) P_{s}}{\zeta} $.
                \If{$ P_{\rm in}^{M}  > \frac{P_{s}}{\zeta}$}
                \State Set $M \leftarrow M + 1$.
                \EndIf
          \EndWhile
  \end{algorithmic}
\end{algorithm}

%\vspace{-1.5em}
\section{Numerical Results}
\label{simul}
%\vspace{-0.5em}

%In this section, we present numerical results to confirm our analysis.
%In this section, we demonstrate the performance of the two proposed schemes: (i) the jointly optimal Tx power allocation and Rx power splitting scheme,
%and (ii) the jointly optimal Tx power allocation and Rx time switching scheme.

In this section, we compare the performance of the proposed and existing schemes.
For the comparison purpose, we extend the existing Tx power allocation and Rx \textit{time switching} scheme of \cite{Zhang13} developed for a single linear EH circuit model, to the adopted model of multiple nonlinear EH circuits, which can be obtained as a special case of Theorem \ref{thm_1} with $\rho = 0$ (i.e., the solution to (P1$'$) when $\rho = 0$).
In the numerical simulations, we consider the Rician fading model: $ g = \sqrt{\frac{r}{r+1}} g_{\rm LOS} + \sqrt{\frac{1}{r+1}} g_{\rm scatter} $,
where $ g $ is the fading channel such that $h = | g |^{2}$; $g_{\rm LOS}$ the line-of-sight (LOS) component; $g_{\rm scatter}$ the scattering component following the Gaussian distribution with zero-mean and variance $\sigma_{\rm scatter}^{2}$; and $r$ the Rician factor. 
We set $r = 2$ and $| g_{\rm LOS} |^{2} =  \sigma_{\rm scatter}^{2} = - 30$ dBW.
%Also, we set $\zeta = 1$, $T = 1$ s, $ P = 2 $ W, and $P_{s} = 0.4 h P$.
Also, we set $T = 1$ s, $\zeta = 1$, $P_{s} = 0.4 \bar{h} P $, $P_{c} = 0.3 P_{s}$, $M_{\max} = \left \lceil \frac{\zeta h P}{ P_{s}} \right \rceil$, and $\sigma_{\rm A}^{2} = \sigma_{\rm cov}^{2} = \sigma^{2}$, 
where $\bar{h} = \frac{r}{r+1} | g_{\rm LOS} |^{2} + \frac{1}{r+1} \sigma_{\rm scatter}^{2}$ is the average channel power gain and $\lceil x \rceil$ is the smallest integer not less than $x$.
The value of $ \sigma^{2} $ is chosen such that $\frac{ \bar{h} P }{ \sigma^{2} } = 20$ dB. The results are averaged over $10^{4}$ different channel realizations.

%\footnote{
%To the best of our knowledge, in the literature, the jointly optimal Tx power allocation and Rx time switching scheme (i.e., the optimal solution to (P1$'$) with $\rho = 0$) has not been developed for neither the linear EH nor the nonlinear EH.
%}

%As a special case of Theorem \ref{thm_1} with $\rho = 0$, it can be shown that the jointly optimal Tx power allocation and Rx \textit{time switching} scheme with $M$ EH nonlinear circuits
%(i.e., the solution to (P1$'$) when $\rho = 0$) is given by
%\begin{align}
%\label{P1_sol_alpha_}
%& \alpha^{*}   = \arg \underset{ \alpha_{\rm low} \leq \alpha \leq 1 }{\max} ( 1 - \alpha) \mathbb{R} \big( P_{\rm ID} (\alpha) , 0 \big),  \\
%\label{P1_sol_pe_}
%& P_{\rm EH}^{*} ( \alpha^{*}  )  =  \begin{cases} \frac{Q + (1 - \alpha^{*}) P_{c} T}{ \alpha^{*} \zeta h T }  , & {\rm if~} \alpha^{*} > 0 \\ 0, & {\rm if~} \alpha^{*} = 0 \end{cases}  , \\
%\label{P1_sol_pi_}
%& P_{\rm ID}^{*} ( \alpha^{*}  )  =  \begin{cases} \frac{\zeta h P T - Q - (1 - \alpha^{*}) P_{c} T}{ (1 -  \alpha^{*} ) \zeta  h T } , &  {\rm if~} \alpha^{*} < 1 \\  0 , & {\rm if~} \alpha^{*} = 1 \end{cases} .
%\end{align}
%In (\ref{P1_sol_alpha}),
%$ \alpha_{\rm low} = \max \{ \alpha_{1} , \alpha_{2}, 0 \} $ and $P_{\rm ID} (\alpha)$ is similarly defined as in (\ref{P1_sol_pi_}),
%where $ \alpha_{1} = 1 -  \frac{ \zeta h P T - Q }{P_{c} T} $ and $ \alpha_{2} = \frac{ Q + P_{c} T }{ M P_{s} T + P_{c} T } $.
%Also, the value of $\alpha^{*} $ can be determined via the one-dimensional searching.

\begin{figure}[!t]
\centering
{
    \includegraphics[width=0.5\textwidth]{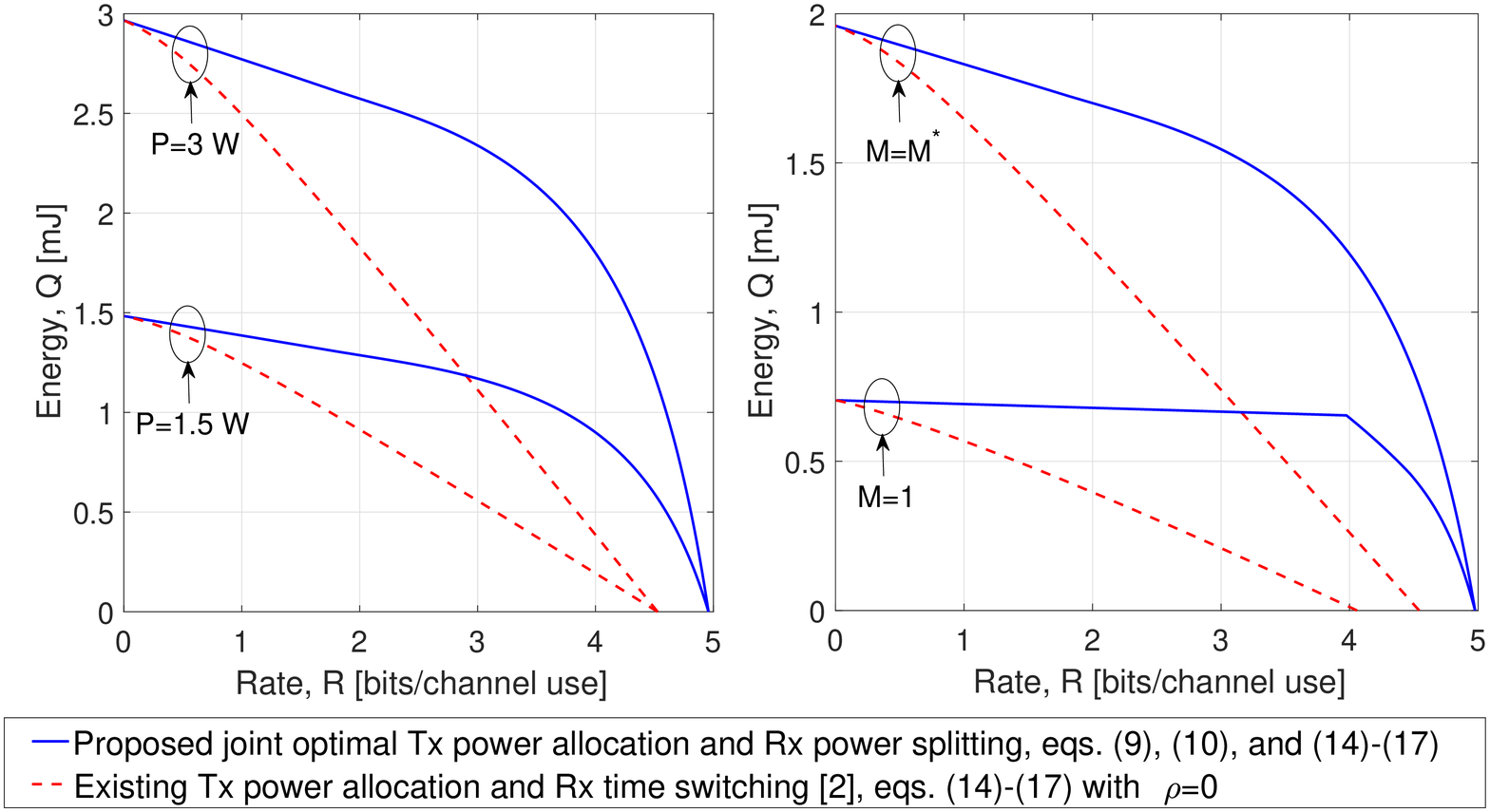}
    \caption{
    R-E tradeoffs of the proposed and existing schemes for $M = M^{*}$ when $P \in \{ 1.5 , 3 \}$ W and for $P = 2$ W when $M = \{ 1, M^{*} \}$.
    }
    \label{fig1}
}
\end{figure}

%In the numerical simulations, we consider a realistic system setting as considered in \cite{Kim16}--\cite{Kang}.
%Specifically, the channel power gain is modeled as $  h  =  1 - \exp \left(   -  \frac{ a_{t} a_{r} }{ (c/f_{c} )^{2} d^{2} }  \right) $ \cite{Huang15},
%where
%$ a_{t} $ is the aperture of the transmit antenna;
%$ a_{r} $ the aperture of the receive antenna;
%$ c $ the speed of light;
%and $ d $ the distance between the transmitter and the receiver.
%Assuming that the receiver is a small sensor,
%we set $ a_{t} = 0.5 $ m, $ a_{r} = 0.01 $ m, $f_{c} = 2.4$ GHz, and $d = 13$ m.

In Fig. \ref{fig1}, the rate-energy (R-E) tradeoffs of the proposed and existing schemes are shown for $M = M^{*}$ when $P \in \{ 1.5 , 3 \}$ W and for $P = 2$ W when $M = \{ 1, M^{*} \}$,
where $M^{*}$ denotes the number of EH circuits obtained by Algorithm \ref{A1}.
Also, the R-E region is defined as $\mathcal{C}_{\texttt{NL}}^{M} = \bigcup_{ \mathbf{P}, \boldsymbol{\rho} } \big\{  (R,Q):   R  \leq  \mathcal{R} ( \mathbf{P}, \boldsymbol{\rho} ) , Q  \leq  \mathcal{Q}_{\texttt{NL}}  ( \mathbf{P}, \boldsymbol{\rho} ) \big\}$, which contains all possible pairs of the rate and harvested energy with $M$ EH nonlinear circuits.
From Fig. \ref{fig1}, it can be seen that the proposed joint Tx power allocation and Rx power splitting scheme considerably outperforms the existing Tx power allocation and Rx time switching scheme.
Also, it can be observed that by adaptively determining the number of EH circuits, the R-E tradeoff performance with nonlinear EH is substantially improved.
This clearly shows the benefit and effectiveness of using multiple EH circuits for the practical SWIPT system.

%From Fig. \ref{fig1}, it can be seen that the proposed scheme significantly outperforms the existing scheme.
%When the number $M$ of EH circuits increases, the performance of the proposed schemes is substantially improved,
%%the rate or R-E tradeoff performance is substantially improved.
%which clearly show the benefit of using multiple EH circuits for the practical SWIPT system with nonlinear EH.
%%Also, the joint Tx power allocation and Rx power splitting scheme considerably outperforms the joint Tx power allocation and Rx time switching scheme.

%\begin{figure}[!t]
%\centering
%{
%    \includegraphics[width=0.35\textwidth]{figure/fig1.eps}
%    \caption{
%    R-E regions for $M \in \{1,2,3 \}$ when $P_{c} > 0$.
%    }
%    \label{fig1}
%}
%\end{figure}

%\begin{figure}[!t]
%\centering
%{
%    \includegraphics[width=0.35\textwidth]{figure/fig2.eps}
%    \caption{
%    R-E regions for $M \in \{1,2,3 \}$ when $P_{c} = 0$.
%    }
%    \label{fig2}
%}
%\end{figure}

%\vspace{-1.5em}
\section{Conclusion}
%\vspace{-0.5em}

We studied the joint Tx power allocation and Rx power splitting for the SWIPT system with nonlinear EH.
We proposed to use multiple EH circuits to overcome the saturation nonlinearity.
Using the realistic nonlinear EH model, we developed the jointly optimal Tx power allocation and Rx power splitting scheme.
%and, as special case of this scheme, we presented the jointly optimal Tx power allocation and Rx time switching scheme.
Also, we developed the algorithm to determine the number of EH circuits.
The obtained results gave us the useful and interesting insights.
The numerical results showed that the SWIPT performance considerably improves when multiple EH circuits are used.

%\appendices

%\numberwithin{equation}{section}

%\setlength{\parskip}{0em}
%\setlength{\parindent}{1em}
%\setlength{\textfloatsep}{5pt}
%\setlength{\abovedisplayskip}{0.1em}
%\setlength{\belowdisplayskip}{0.1em}
%\setlength{\topskip}{0em}
%%\setlength{\columnsep}{0em}
%%\setlength{\topmargin}{0em}
%\setlength{\abovecaptionskip}{0em}
%\setlength{\belowcaptionskip}{0em}

%\textheight 59pc

%\vspace{-1.em}
\section*{Appendix A: Proof of Lemma \ref{lem_1}}
%\vspace{-0.5em}
\renewcommand\theequation{A.\arabic{equation}}

The optimal objective value of (P1) depends only on the cardinality of the set $\Omega$.
Also, in (P1), it must be $| \Omega^{C} | = N - | \Omega | = \alpha N$ for some $0 \leq \alpha \leq 1$
since the amount of circuit power consumption reduces from $P_{c}$ to $(1-\alpha) P_{c}$.
Thus, without loss of any optimality, we can take $\Omega = \{ \alpha N + 1 , \cdots, N \}$ and $\rho_{k} = 1$, $k \in \Omega^{C} = \{ 1,\cdots, \alpha N \} $.
Consequently, given $\alpha$ and $\boldsymbol{\rho}$, the optimization of (P1) becomes convex in $\mathbf{P}$.
From the Karush-Kuhn-Tucker conditions, we have $P^{*} (\alpha, \rho_{k}) = \frac{1}{\nu - \mu \zeta \rho_{k}  h} - \frac{1}{ h} \left( \sigma_{ {\rm A}}^{2} + \frac{\sigma_{ {\rm cov} }^{2}}{1 - \rho_{k}} \right) $, $k=\alpha N + 1,\cdots, N$, where $\mu \geq 0$ and $\nu \geq 0$ are the Lagrange multipliers associated with the constraints of (\ref{P1_const_1}) satisfying $\mu - \nu \zeta  h \geq \frac{h}{\sigma_{ {\rm A} }^{2} + \sigma_{ {\rm cov} }^{2}}$.
Given $\alpha$, both $\mathbb{R} \big( P^{*} (\alpha, \rho_{k}) , \rho_{k} \big)$ and $ \mathbb{Q}_{\texttt{NL}} \left( P^{*} ( \alpha, \rho_{k} ), \frac{\rho_{k}}{M} \right) $ are concave in $\rho_{k}$, $k=\alpha N + 1,\cdots, N$. Therefore, we have $ \frac{1}{N} \sum_{k=\alpha N + 1}^{N} C \big( P^{*} (\alpha, \rho_{k}) , \rho_{k} \big)  \leq  (1 - \alpha)  C \big( P^{*} (\alpha, \rho), \rho \big) $
and $ \frac{1}{N}  \sum_{k=\alpha N + 1}^{N}   E_{\texttt{NL}} \left( P^{*} ( \alpha, \rho_{k} ), \frac{\rho_{k}}{M} \right)  \leq (1 - \alpha)  E_{\texttt{NL}} \left( P^{*} ( \alpha, \rho ), \frac{\rho}{M} \right) $,
where $\rho = \frac{ 1} {(1 - \alpha) N}  \sum_{k=\alpha N + 1}^{N} \rho_{k} $.
From this, with given $\alpha$, the solution to (P1) can be written as in (\ref{P_k}) and (\ref{rho_k}),
where $P_{\rm EH} =  \frac{1}{N}  \sum_{k=1}^{\alpha N} P_{k}$ and $P_{\rm ID} = P^{*} (\alpha, \rho)$, $k=\alpha N+1, \cdots, N$.

%\vspace{-1.em}
\section*{Appendix B: Proof of Theorem \ref{thm_1}}
%\vspace{-0.5em}
\renewcommand\theequation{B.\arabic{equation}}

In (P1$'$), both the constraints of (\ref{P1_const_1}) must be satisfied with equalities.
Thus, the constraint of (\ref{P1_const_1}) can be divided into the following five constraints:
(\ref{P1_const_1}a): $ Q_{\rm EH} + Q_{\rm ID} - (1 - \alpha) P_{c} T = Q $; (\ref{P1_const_1}b): $ \alpha \zeta h P_{\rm EH} T \geq Q_{\rm EH} $; (\ref{P1_const_1}c): $ (1 - \alpha)  \zeta \rho h P_{\rm ID} T \geq Q_{\rm ID} $;
(\ref{P1_const_1}d): $ \zeta h P_{\rm EH} \leq M P_{s} T $; and (\ref{P1_const_1}e): $ \zeta \rho h P_{\rm ID} \leq M P_{s} T $.
%First, we consider the case when $P_{c} > 0$.
If $ \alpha = 0 $, we have $ P_{\rm EH} = 0 $, $P_{\rm ID} = P$, and $ \rho = \frac{ Q  }{ \zeta h P T } $.
If $\alpha = 1$, we can set $P_{\rm ID} = 0$.
If $0 < \alpha < 1$, from (\ref{P1_const_1}b) and (\ref{P1_const_1}c),
we have $P_{\rm EH} = \frac{ Q_{\rm EH} }{ \alpha \zeta h T }$ and $P_{\rm ID} = \frac{ Q_{\rm ID} }{ (1 - \alpha) \zeta \rho h T }$, respectively.
Thus, from (\ref{P1_const_1}a),
we obtain $Q_{\rm EH} = \frac{1}{1 - \rho} \left(  Q + (1 - \alpha) P_{c} T - \zeta \rho h P T \right)$
and $Q_{\rm ID} = \frac{\zeta \rho h }{1 - \rho} \left(  P T - Q - (1 - \alpha) P_{c} T \right)$, respectively.
Substituting these $Q_{\rm EH}$ and $Q_{\rm ID}$ into $P_{\rm EH}$ and $P_{\rm ID}$, respectively,
we can obtain $P_{\rm EH} (\alpha)$ and $P_{\rm ID} (\alpha)$ similarly as in (\ref{P1_sol_pe}) and (\ref{P1_sol_pi}), respectively.
To satisfy (\ref{P1_const_1}d), it must be $ \rho \geq \max \left\{  \frac{ Q + (1-\alpha) P_{c} T - \alpha M P_{s} T }{  \zeta h P T - \alpha M P_{s} T }, 0 \right\} $.
Also, to satisfy both (\ref{P1_const_1}e) and $P_{\rm EH} (\alpha) \geq 0$,
it must be $ \rho \leq \min \{ \rho_{1} (\alpha) , \rho_{2} (\alpha) \} $.
To ensure $\rho_{1} (\alpha) \leq 1$, $\rho_{2} (\alpha) \leq 1$, and $P_{\rm ID}(\alpha) \geq 0$,
it must be $ \alpha_{\rm low} \leq \alpha \leq 1$.
Since the objective function is increasing in $\rho$ when $P_{\rm ID} (\alpha)$ is substituted, the optimal $\rho$ is given by $\rho(\alpha) = \min \{ \rho_{1} (\alpha) , \rho_{2} (\alpha) \} $.
Then the optimal $\alpha$ can be determined by maximizing $( 1 - \alpha) \mathbb{R} \big( P_{\rm ID} (\alpha) , \rho (\alpha) \big)$ over $ \alpha_{\rm low} \leq \alpha \leq 1 $.
%Second, we consider the case when $P_{c} = 0$.
%In this case, $( 1 - \alpha) C ( P_{\rm ID} (\alpha) , \rho (\alpha) )$ is monotonically decreasing in $\alpha$.
%Thus, we have $\alpha = 0$ and $P_{\rm EH} = 0$. Also, following the above procedures, we have $\rho = \frac{Q}{\zeta h P T}$ and $P_{\rm ID} = P$.
%%Thus, the results of (\ref{P1_sol_alpha})--(\ref{P1_sol_pi}) follow.

%\vspace{-1.em}

%\section*{Acknowledgment}


\ifCLASSOPTIONcaptionsoff
  \newpage
\fi

%\begin{biography}[{\includegraphics[width=1in,height=1.25in,clip,keepaspectratio]{mshell}}]{Michael Shell}

%\end{IEEEbiography}


%\newpage



%\vfill

\end{document}